\documentclass[twoside,twocolumn,9pt]{article}
\usepackage{extsizes}

\usepackage[super,sort&compress,comma]{natbib} 
\usepackage[version=3]{mhchem}
\usepackage[left=1.5cm, right=1.5cm, top=1.785cm, bottom=2.0cm]{geometry}
\usepackage{balance}
\usepackage{times,mathptmx}
\usepackage{sectsty}
\usepackage{graphicx} 
\usepackage{lastpage}
\usepackage[format=plain,justification=justified,singlelinecheck=false,font={stretch=1.125,small,sf},labelfont=bf,labelsep=space]{caption}
\usepackage{float}
\usepackage{fancyhdr}
\usepackage{fnpos}
\usepackage[english]{babel}
\addto{\captionsenglish}{%
  
}
\usepackage{array}
\usepackage{droidsans}
\usepackage{charter}
\usepackage[T1]{fontenc}
\usepackage[usenames,dvipsnames]{xcolor}
\usepackage{setspace}
\usepackage[compact]{titlesec}
\usepackage{hyperref}

\usepackage[normalem]{ulem}
\usepackage{siunitx}
\usepackage{todonotes}

\definecolor{cream}{RGB}{222,217,201}

\begin{document}

\pagestyle{fancy}
\thispagestyle{plain}
\fancypagestyle{plain}{

\renewcommand{\headrulewidth}{0pt}
}

\makeFNbottom
\makeatletter
\renewcommand\LARGE{\@setfontsize\LARGE{15pt}{17}}
\renewcommand\Large{\@setfontsize\Large{12pt}{14}}
\renewcommand\large{\@setfontsize\large{10pt}{12}}
\renewcommand\footnotesize{\@setfontsize\footnotesize{7pt}{10}}
\makeatother

\renewcommand{\thefootnote}{\fnsymbol{footnote}}
\renewcommand\footnoterule{\vspace*{1pt}%
\color{cream}\hrule width 3.5in height 0.4pt \color{black}\vspace*{5pt}} 
\setcounter{secnumdepth}{5}

\makeatletter 
\renewcommand\@biblabel[1]{#1}            
\renewcommand\@makefntext[1]%
{\noindent\makebox[0pt][r]{\@thefnmark\,}#1}
\makeatother 
\renewcommand{\figurename}{\small{Fig.}~}
\sectionfont{\sffamily\Large}
\subsectionfont{\normalsize}
\subsubsectionfont{\bf}
\setstretch{1.125} 
\setlength{\skip\footins}{0.8cm}
\setlength{\footnotesep}{0.25cm}
\setlength{\jot}{10pt}
\titlespacing*{\section}{0pt}{4pt}{4pt}
\titlespacing*{\subsection}{0pt}{15pt}{1pt}

\fancyfoot{}
\fancyfoot[LO,RE]{\vspace{-7.1pt}\includegraphics[height=9pt]{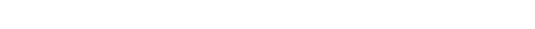}}
\fancyfoot[CO]{\vspace{-7.1pt}\hspace{13.2cm}\includegraphics{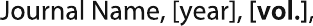}}
\fancyfoot[CE]{\vspace{-7.2pt}\hspace{-14.2cm}\includegraphics{head_foot/RF}}
\fancyfoot[RO]{\footnotesize{\sffamily{1--\pageref{LastPage} ~\textbar  \hspace{2pt}\thepage}}}
\fancyfoot[LE]{\footnotesize{\sffamily{\thepage~\textbar\hspace{3.45cm} 1--\pageref{LastPage}}}}
\fancyhead{}
\renewcommand{\headrulewidth}{0pt} 
\renewcommand{\footrulewidth}{0pt}
\setlength{\arrayrulewidth}{1pt}
\setlength{\columnsep}{6.5mm}
\setlength\bibsep{1pt}

\makeatletter 
\newlength{\figrulesep} 
\setlength{\figrulesep}{0.5\textfloatsep} 

\newcommand{\topfigrule}{\vspace*{-1pt}%
\noindent{\color{cream}\rule[-\figrulesep]{\columnwidth}{1.5pt}} }

\newcommand{\botfigrule}{\vspace*{-2pt}%
\noindent{\color{cream}\rule[\figrulesep]{\columnwidth}{1.5pt}} }

\newcommand{\dblfigrule}{\vspace*{-1pt}%
\noindent{\color{cream}\rule[-\figrulesep]{\textwidth}{1.5pt}} }

\makeatother

\twocolumn[
  \begin{@twocolumnfalse}
  {\includegraphics[height=30pt]{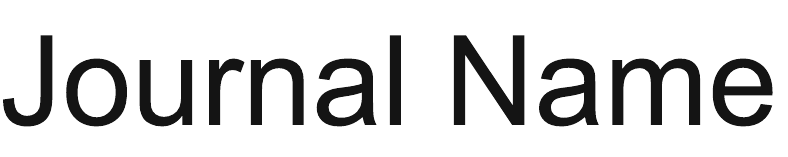}\hfill%
   \raisebox{0pt}[0pt][0pt]{\includegraphics[height=55pt]{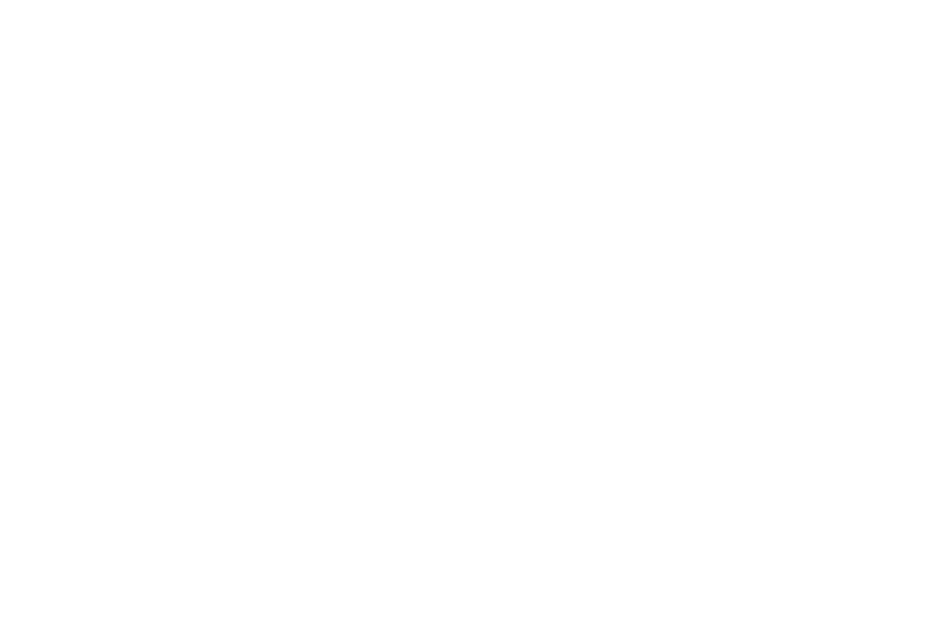}}%
   \\[1ex]%
   \includegraphics[width=18.5cm]{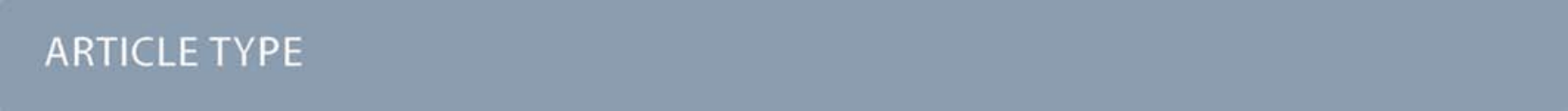}
  }\par
  \vspace{1em}
\sffamily
\begin{tabular}{m{4.5cm} p{13.5cm} }

  \includegraphics{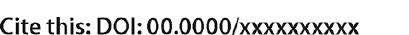}
  & \noindent\LARGE{\textbf{Frequency-dependent magnetic
    susceptibility of magnetic nanoparticles in a polymer
    solution: a simulation study}} \\
  \vspace{0.3cm} & \vspace{0.3cm} \\

  & \noindent\large{Patrick Kreissl,$^{\ast}$\textit{$^{a}$} Christian Holm\textit{$^{a}$} and Rudolf Weeber\textit{$^{a}$}} \\

  \includegraphics{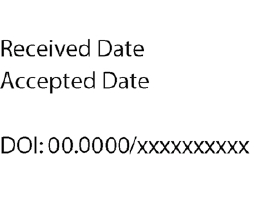} & \noindent\normalsize{
                                      Magnetic composite materials
                                      \textit{i.e.}\ elastomers,
                                      polymer gels, or polymer solutions
                                      with embedded
                                      magnetic nanoparticles are
                                      useful for many technical
                                      and bio-medical
                                      applications. However, the microscopic
                                      details of the coupling
                                      mechanisms between the magnetic
                                      properties of the particles and
                                      the mechanical properties of the
                                      (visco)elastic polymer matrix
                                      remain unresolved. Here we study
                                      the response of a
                                      single-domain spherical magnetic
                                      nanoparticle that is suspended in a
                                      polymer solution to alternating
                                      magnetic fields. As
                                      interactions we consider only excluded volume
                                      interactions with the polymers
                                      and hydrodynamic interactions
                                      mediated through the solvent. The AC
                                      susceptibility spectra are
                                      calculated using a
                                      linear response Green-Kubo
                                      approach, and the
                                      influences of changing polymer
                                      concentration and polymer
                                      length are investigated. Our
                                      data is compared to
                                      recent measurements of the AC susceptibility
                                      for a typical magnetic
                                      composite system [Roeben
                                      \textit{et al., Colloid and
                                      Polymer Science}, 2014,
                                      2013--2023], and demonstrates the
                                      importance of hydrodynamic
                                      coupling in such systems.}
  \\

\end{tabular}

 \end{@twocolumnfalse} \vspace{0.6cm}

  ]

\renewcommand*\rmdefault{bch}\normalfont\upshape
\rmfamily
\section*{}
\vspace{-1cm}


\footnotetext{\textit{$^{a}$~Institute for Computational Physics,
    University of Stuttgart, Allmandring 3, 70569 Stuttgart,
    Germany. Fax: 49 (0)711 68563658; Tel: 49 (0)711 68563593; E-mail:
    pkreissl@, holm@, weeber@icp.uni-stuttgart.de}}




\section{Introduction}

Magnetic nanoparticles (MNPs) have been connected to many technical
applications and are also promising candidates for novel bio-medical
treatments.  They can be manufactured by splitting larger magnetic
objects into smaller particles (\textit{e.g.}~by grinding) or
synthesized using chemical processes, allowing for particles of
well-defined shape, size, and material properties.\cite{laurent08a,
  lopezperez97a, faraji10a, roeben14a, remmer17a, sun04a, sen15a,
  gunther11a, markert11a} As the human body mass is transparent
to static magnetic fields, MNPs that are injected into
a body can be moved by application of inhomogeneous external
fields. This allows an externally controlled accumulation of MNPs in
specific places, \textit{e.g.}~in tumors.  Loading the particles with
therapeutics before injection, they can be used as a targeted drug
delivery system -- an important step toward locally constrained
chemotherapy.\cite{tietze15a, tietze13a, goodwin99a, alexiou05a,
  alexiou07a} Another promising approach to cancer treatment based on
MNPs is the local heating of cancer cells, so-called
`hyperthermia'.\cite{perigo15a, engelmann18a, hergt06a, aqil08a,
  lao04a, huang13a} Typical bio-medical applications of MNPs will be
situated in a polymer-rich, \textit{i.e.}, complex cellular
environment.  The use of MNPs in a technical context has mainly two
aspects. On one hand, they can be embedded in a polymeric gel or
suspension to create a soft composite material. The coupling between
the magnetic properties of the particles and (visco)elastic
properties of the polymer matrix allows for the dynamic control of
some aspects of the composite. This includes, \textit{e.g.}, changing
their shape, motion, or elastic properties using external fields,
which makes them interesting candidates for a variety of
applications.\cite{gollwitzer08a, szabo98a, ramanujan06a, odenbach16a,
  mitsumata11a, monz08a, varga06a, volkova17a} On the other hand, MNPs
can be viewed as probes that are used to asses the local behaviour of a
polymeric environment.  A well established experimental technique for
that is the so-called \emph{AC susceptometry}, where a (sinusoidal
modulated) external magnetic field is applied to a
sample. Frequency-dependent susceptibility spectra are obtained by
simultaneous measurement of the probes' magnetization
response. Besides the insight gained into the magnetic behaviour of the
composite as a whole, there exist a number of theoretical models that
connect the susceptibility to frequency-dependent mechanical
properties. Using, \textit{e.g.}, extended Debye
models,\cite{ludwig10a, caleroddelc11a, feyen08a} the
Gemant-DiMarzio-Bishop model,\cite{dimarzio74a, gemant35a, ilg18a,
  havriliak95a, diazcalleja05a, niss05a} or the theoretical model
derived by \citet{raikher01b}, one may infer the
frequency-dependent local viscosity and even elastic moduli on the
particle scale, a method dubbed \emph{magnetic nanorheology}.

A good understanding of the coupling mechanisms is key, when it comes
to tailoring the properties of MNPs to a specific bio-medical and
technical use case.  Unfortunately, often the details of the coupling
between the particles and the polymeric environment are not fully
known and moreover difficult to (directly) determine
experimentally. Here, computer simulations are a powerful tool,
allowing the study of well-defined model systems. Aspects of the
computational models can be easily modified, where it might be
difficult to do so experimentally.  Computational models exist on
different scales of modelling.  While continuum descriptions are
well-suited to model MNP--polymer composites on a
macro-level,\cite{stolbov11a, kalina16a, metsch16a, menzel17a,
  puljiz19a} they typically lack details on the length scale of single
polymers.  This, however, will be necessary to study how individual
changes to the polymer matrix affect its coupling to the embedded
MNPs.  In this work, we therefore present a particle-based molecular
dynamics (MD) simulation model for MNPs immersed in a polymer
solution.  As explicit resolution of the polymers comes at a high
computational cost -- in particular at high concentrations and large
sample sizes -- we focus on the efficient modelling of such systems.
The model, which will be presented in section~\ref{sec:model}, uses
coarse-grained polymers, that are coupled to a computationally
efficient lattice-Boltzmann hydrodynamics solver; MNPs are modelled as
so-called `raspberry particles',\cite{lobaskin04a, fischer15a} in
order to couple translations as well as rotations hydrodynamically to
the environment. 
We use our model to simulate a typical nanorheological system. As a
representative example, we use the system of \citet{roeben14a}
throughout this publication. The parameters required for simulation
using our model are specified in section~\ref{sec:parameters}.  To
compare to the experiment, we need to obtain AC susceptibility spectra
from the simulations, which can be done in two ways.  While the
susceptibility can be directly measured for specific frequencies
(section~\ref{sec:direct}), our main method is based on Green-Kubo
linear response theory (section~\ref{sec:gk}). It allows us to obtain
a complete susceptibility spectrum from a single (however more
expensive) simulation.  With this approach, we demonstrate that our
model reproduces the experimental trends. Using the Green-Kubo
approach, we run simulations to illustrate how changes to the polymer
matrix affect the resulting susceptibility. Higher values for both,
the polymer volume fraction and the polymers' chain length, are found
to independently cause a shift of the susceptibility spectra towards
lower frequencies (section~\ref{sec:results}), which is in accordance
with the experimental observations of \citet{roeben14a}. We end with
a discussion of our model and the results obtained our the simulations of the experimental system.

\section{Computational model \label{sec:model}}

In this section we present an efficient MD model for simulations of
typical nanorheological systems, \textit{i.e.} (magnetically blocked)
MNPs immersed in polymer suspension.  Such systems consist of three
main components that have to be included in the model, namely the
polymers, the fluid, and the magnetic particles.

In MD simulations, parts of the system are represented by interacting
MD particles, which are propagated through space and time by
time-discrete integration of Newton's equation of motion.  For the
polymers, the so-called bead--spring formalism is used.  Every polymer
chain is explicitly resolved using MD beads, which are inter-connected
by an attractive potential -- here, we use a simple harmonic
interaction of the form
\begin{equation}
  \varphi_\mathrm{harm} = \frac{1}{2} k (r - r_0)^2,
\end{equation}
where $r$ is the distance between two consecutive MD beads within a
polymer chain, $r_0$ is the potential's equilibrium distance and the
spring constant $k$ determines the stiffness of the bond.  The MD
beads themselves interact \textit{via} a purely repulsive
Weeks-Chandler-Andersen (WCA) potential\cite{weeks71a}
\begin{equation}
\varphi_\mathrm{WCA}(r) =
\begin{cases}
  \varphi_\mathrm{LJ} + \epsilon & \mathrm{if}\ r < 2^{1/6} \sigma, \\
  0 & \mathrm{otherwise,} \label{eqn:wca}
\end{cases}
\end{equation}
based on the well-known 6-12 Lennard-Jones (LJ) pair potential,
\begin{equation}
  \varphi_\mathrm{LJ} = 4 \epsilon \left[ \left( \frac{\sigma}{r} \right)^{12} - \left( \frac{\sigma}{r} \right)^{6} \right], \label{eqn:lj}
\end{equation}
with $\epsilon$ the depth of the LJ potential well and $\sigma$ the
distance at which $\varphi_\mathrm{LJ} = 0$.  The repulsive
interaction effectively models excluded volume.  Note that typically
every MD bead within the polymer will represent more than one
`real-world' monomer unit. This \emph{coarse-graining} approach allows
for significantly larger system sizes or simulation times as it
reduces the number of interacting particles.

To include the hydrodynamic coupling between MNPs and close-by
polymers in our model, hydrodynamics are solved using the
lattice-Boltzmann (LB) algorithm.\cite{mcnamara88a, krueger17a} This
very efficient grid-based hydrodynamics solver can be
straight-forwardly parallelized, has good scaling behaviour, and there
also exist fast implementations using graphics cards. Their
computational performance make LB algorithms well-suited for the
mesoscale. On the small length-scales of our system, thermal
fluctuations are an important part of modelling. The LB fluid can be
thermalized by adding stochastic fluctuations to the stress tensor,
which guarantees local mass and momentum conservation.\cite{ladd94a,
  dunweg09a, landau59a} To couple MD particles to this fluid, a
dissipative friction force is used, which also has to be
thermalized. Following \citet{ahlrichs99a}, the coupling force is
given by
\begin{equation}
  \vec{F} = -\gamma (\vec{u}_\mathrm{fluid} - \vec{u}_\mathrm{part}) + \vec{\mathcal{F}}. \label{eqn:coupling_force}
\end{equation}
Its strength is determined by the friction parameter $\gamma$ and the
difference of fluid velocity $\vec{u}_\mathbf{fluid}$ and velocity of
the particle $\vec{u}_\mathrm{part}$. Fluctuations are incorporated
\textit{via} the stochastic force $\vec{\mathcal{F}}$ of zero mean and
\begin{equation}
  \langle \mathcal{F}_{i}(t) \mathcal{F}_j (t') \rangle = 2 \gamma k_\mathbf{B} T \delta_{ij} \delta(t-t') \label{eqn:harmonic},
\end{equation}
where $\delta$ is the Kronecker delta.

\begin{figure}[h]
	\centering
	\includegraphics[width=0.50\linewidth]{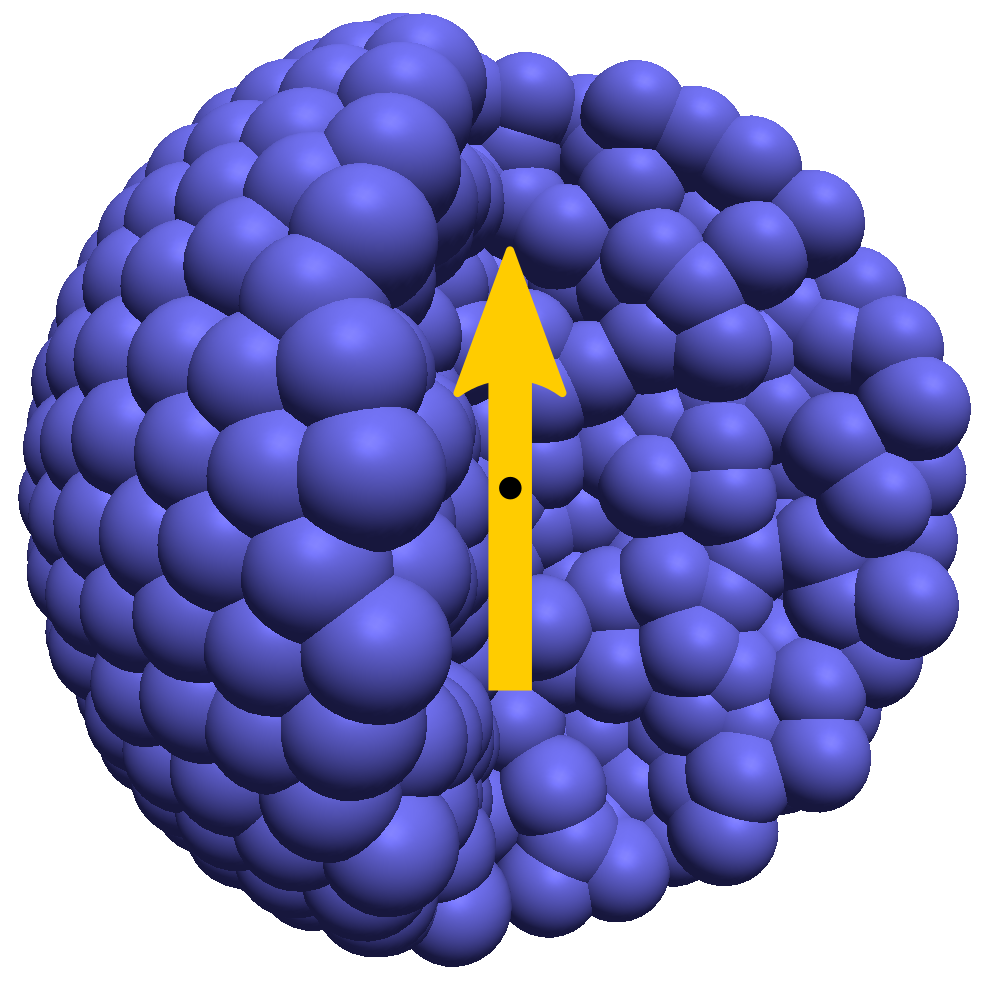}
	\caption{Cut through the spherical raspberry particle used for
          this publication. It consists of homogeneously distributed
          fluid coupling points (blue) forming a rigid body. Newton's
          equation of motion is integrated only for its centre
          (black), which carries the particle's mass, as well as its
          magnetic moment (indicated by the yellow arrow).}
	\label{fgr:raspberry_cut}
\end{figure}

In typical experimental setups, the MNPs are at least one order of
magnitude larger than the polymers' persistence length
$\lambda_\mathrm{p}$ (\textit{e.g.}, in the experimental system which
serves as a basis for the parameter choices in
section~\ref{sec:parameters},\cite{roeben14a}
$r_\mathrm{h} / \lambda_\mathrm{p} \approx 18.9$, with hydrodynamic
radius $r_\mathrm{h}$ of the magnetic particle).  To properly capture
the hydrodynamic behaviour of the MNPs in the model, it does not
suffice to just use a single coupling point per particle.  With such
an approach every MNP would couple to only a single lattice site per
time step, with no rotational coupling. In addition, the hydrodynamic
impact of the MNP's shape would be neglected.  Thus, to obtain the
physically expected hydrodynamic behaviour, the so-called `raspberry
model' is used for the MNPs, which means homogeneously filling their
shape with fluid coupling points that are rigidly bound to a central
MD bead.\cite{lobaskin04a} For preparation of the raspberry we follow
the procedure described by \citet{fischer15a}, which we will only
briefly sketch here. For in-depth information please consult the
referenced publication.  Creating the raspberry particle happens in a
two-step process: resolving the surface and then homogeneously filling
its volume, which is necessary to obtain consistent translational and
rotational diffusivities $D_\mathrm{t}$ and
$D_\mathrm{r}$.\cite{fischer15a, ollila13a}  The preparation process
uses a MD approach. To resolve the surface of the raspberry, (mobile)
MD beads are put into an empty simulation box. A harmonic potential
(see eq.~\ref{eqn:harmonic}) that is shifted by the desired raspberry
radius $r_0 = r_\mathrm{rasp}$, is used to force the beads onto a
spherical shell.  To homogenize the surface density, a purely
repulsive WCA potential (see eq.~\ref{eqn:wca}) is added between the
MD beads. By slowly increasing the harmonic spring constant $k$ to an
extremely high target value, MD beads end up forming a homogeneous
surface shell. After correcting small deviations of the beads' desired
distances from the raspberry centre ($|\vec{r}|=r_\mathrm{rasp}$),
they are rigidly connected to the centre position and the harmonic
bonds removed. Now the raspberry can be filled by adding MD beads
inside the hollow shell, which interact again \textit{via} a repulsive
WCA potential. To prevent numerical instabilities, the resulting
forces are initially capped. The positions of the MD beads are allowed
to evolve using MD (with a Langevin thermostat) while slowly raising
the force cap. This procedure typically leads to a homogeneous
distribution of MD beads, which is tested by tracking the shell-wise
particle density, as well as the geometric centre of all raspberry
bead positions (it should coincide with the raspberry centre). Once
these conditions are fulfilled, the MD beads are rigidly bound to the
raspberry centre.\cite{fischer15a} All beads of the resulting
raspberry-like shaped rigid body now act as fluid coupling
points. They couple to the LB fluid \textit{via} the same friction force
as the other MD beads (see eq.~\ref{eqn:coupling_force}). The equation
of motion, however, is only integrated for the body's central bead,
which carries the mass $M_\mathrm{rasp}$ and moment of inertia
$I = 2 M_\mathrm{rasp} r_\mathrm{rasp}^2 / 5$ of the particle, as well
as its magnetic moment $m$. Figure~\ref{fgr:raspberry_cut} shows a cut
through the raspberry particle used for this publication.
Imposing a meaningful potential between the central bead and polymer
MD particles, one can model respective interactions -- as we will see
in section \ref{sec:parameters}, we choose a purely repulsive WCA
potential, here, but more involved interaction potentials would also
be possible.  Note that for a spherical particle these interactions
can only lead to translations. Coupling of the rotational behaviour
happens only \textit{via} hydrodynamics which are thus essential to
our model.

\section{Model parameters \label{sec:parameters}}

To establish a close relationship between experiment and modelling, we
set up simulations based on the experimental system used by
\citet{roeben14a} and \citet{hess19a} for nanorheological
measurements. It consists of spherical, magnetically blocked
single-domain cobalt ferrite (\ce{CoFe2O4}) nanoparticles immersed in
an aqueous polyethylene glycol (PEG) solution.  The \ce{CoFe2O4}
particles act as probes for investigating the nanorheological
properties of the polymer solution.  Their magnetic moment is small,
and they are used at very high dilutions (\SI{3.6e-2}{\percent} volume
fraction\cite{roeben14a}).  Therefore, dipolar interactions between
the MNPs do not play a major role (for an estimate of their relative
strength see the calculation of $\lambda$ in
section~\ref{sec:results}).  Consequently, we neglect dipolar
interactions in our model.  This allows us to simulate only a portion
of the system containing a single magnetic particle.  Periodic
boundary conditions are used to eliminate boundary effects, but with a
very large simulation box to minimize hydrodynamic interactions
between the MNP and its periodic images.

All simulations for this publication were performed using the MD
software package \emph{ESPResSo}.\cite{weik19a} This software uses the
concept of simulation units, that allows the user to freely choose the
mass, length, and energy scales of the system, which in combination
determine the simulated time scale. For reference, the values of
important quantities describing the system are summarized in
table~\ref{tbl:parameters} using both, SI and simulation units,
respectively.

The length scale in our model is prescribed by the hydrodynamic radius
of the raspberry particle $r_\mathrm{rasp}$, which corresponds to the
experimental value of
$r_\mathrm{h} = \SI{7.2}{\nano\meter}$.\cite{roeben14a} As simulation
domain, we use a cubical box with side length
$L_\mathbf{box} = 10 r_\mathrm{rasp}$.
\begin{table}[h]
\small
\caption{Overview of the parameters we use for the simulation unit
  system, for some characteristic system quantities.}
  \label{tbl:parameters}
  \begin{tabular*}{0.48\textwidth}{@{\extracolsep{\fill}}lll}
    \hline
    Quantity & Value (SI units) & Value (simulation units) \\
    \hline
    particle radius $r_\mathrm{rasp}$ & \SI{7.2}{\nano\meter} & \SI{4}{[x]} \\
    thermal energy & $k_\mathrm{B}$ $\cdot$ \SI{300}{\kelvin} & \SI{1}{[E]} \\
    solvent density $\rho_\mathrm{w}$ & \SI{1e3}{\kg\per\meter\cubed} & \SI{1}{[m]/[x]\cubed} \\
    kinematic viscosity $\nu$ & \SI{8.9e-8}{\meter\squared\per\second} & \SI{1.86}{[x]\squared\per[t]} \\
    mass unit & \SI{5.8e-24}{\kilo\gram} & \SI{1}{[m]} \\
    time unit & \SI{67.4}{\pico\second} & \SI{1}{[t]} \\
    \hline
  \end{tabular*}
\end{table}
The energy scale is given by the thermal energy of the system. As the
experiments are performed at $T_\mathrm{exp} = \SI{25}{\celsius}$, we
use $T = \SI{300}{\kelvin}$ in our model, prescribing the energy scale
$\SI{1}{[E]} = k_\mathrm{B} \cdot \SI{300}{\kelvin}$.

For the fluid, we make the following choices: the size of the LB grid
cells is set to $a_\mathrm{grid} = \SI{1}{[x]} = r_\mathrm{rasp} / 4$
yielding $[x] = \SI{1.8}{\nano\meter}$; LB fluid density is set to
\SI{1}{[m]\per[x]\cubed} corresponding to water density
$\rho_\mathrm{w} = \SI{1e3}{\kilogram\per\meter\cubed}$, thereby also
defining the mass scale $[m] = \SI{5.8e-24}{\kilo\gram}$.  For the
magnetic particle, we use the bulk density of \ce{CoFe2O4},
$\rho_{\ce{CoFe2O4}} =
\SI{5.3e3}{\kilogram\per\meter\cubed}$.\cite{shinde13a} We make a
simplifying assumption concerning the fluid viscosity by choosing its
value as one tenth the experimental kinematic viscosity of water,
$\nu_\mathrm{exp} = 0.1 \nu_\mathrm{w} =
\SI{8.9e-8}{\meter\squared\per\second}$,\cite{kestin78a} in simulation
units $\nu = \SI{1.86}{[x]\squared\per[t]}$.  This effectively
decreases the decay time of the magnetic moment of the MNP through
rotational motion, which reduces simulation time by roughly one order
of magnitude.  Per every MD time step $\Delta t = \SI{0.01}{[t]}$, the
LB fluid field is updated. The simulated time scale is fully
determined by the prescribed length, energy, and mass scales yielding
$[t] = \SI{67.4}{\pico\second}$.

\begin{figure}[h]
	\centering
	\includegraphics[width=\linewidth]{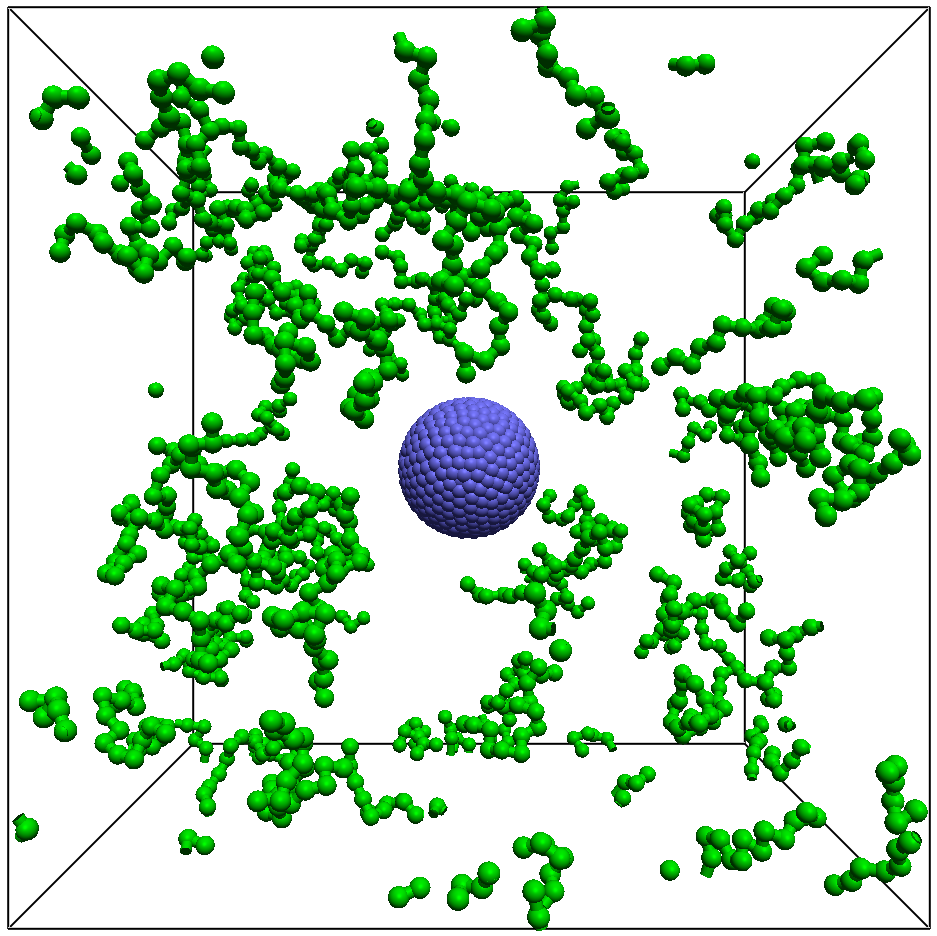}
	\caption{Visualization of the simulation model described in
          section~\ref{sec:model} using the parameters of
          section~\ref{sec:parameters}. Raspberry coupling points are
          colored blue, (coarse-grained) polymers green.  Here, each
          polymer chain consists of 15 beads. For visibility reasons,
          the polymer volume fraction in the shown setup is only
          $\phi = 0.01$. Typical volume fractions used for our
          simulations are significantly higher, up to $\phi = 0.25$.}
	\label{fgr:snapshot}
\end{figure}

In the experiment, PEG is the polymer species being
used. Experimentally, the mean monomer length in PEG is
\SI{2.78}{\angstrom},\cite{oesterhelt99a} the chain's persistence
length in aqueous solution
$\lambda_\mathrm{p} \approx \SI{3.8}{\angstrom}$.\cite{mark65a,
  kienberger00a} To accurately capture the experimental size ratio
between magnetic particles and persistence length of PEG would be
computationally exceedingly demanding, limiting investigations to
small polymer concentrations or simulation times. Our aim is, however,
to end up with a computationally model that can handle polymer volume
fractions up to as high as $\sim \SI{25}{\percent}$. In addition,
simulations of such systems have to capture both, the relaxation time
of the polymer suspension and the relatively long self-diffusion time
of the magnetic particle, making coarse-graining of the polymers
indispensable. While this may change the strength of some effects in
the system, general trends are expected to remain unaffected -- for
results and discussion, please see section~\ref{sec:susceptometry}.
As water is a good solvent for PEG, excluded volume has to be
considered for the polymers, which is accounted for using a purely
repulsive WCA interaction between monomer beads (see
eq.~\ref{eqn:wca}). The diameter of the monomer beads is given by the
choice of the WCA parameter, here
$\sigma_\mathrm{mono} = \SI{1}{[x]}$; we set
$\epsilon_\mathrm{mono} = \SI{1}{[E]}$.  The equilibrium length of the
harmonic potential which connects monomer beads within each polymer
chain (eq.~\ref{eqn:harmonic}) is set to $r_0 = \sigma_\mathrm{mono}$,
the spring constant $k = \SI{200}{[E]\per[x]\squared}$.
Finally, to prevent polymers from penetrating the `solid' MNP, an additional WCA potential is required between the monomer beads of the polymers and the raspberry particle. In principle, this interaction could be set between monomer beads and the coupling points of the raspberry. Since, in general, the coupling points do not smoothly resolve the raspberry surface, any remaining surface roughness introduces additional rotational coupling.
As detailed in section~\ref{sec:model}, using the raspberry model the main concern is to obtain consistent translational and rotational diffusive behaviour for the MNP, which can be challenging in itself and does in no way guarantee a smooth resolution of the MNP's surface.
To avoid artefacts in the rotational coupling, we therefore prescribe the repulsive WCA potential between the monomer beads and the centre or the raspberry particle.
The parameters for this mixed
interaction are $\epsilon_\mathrm{mix} = \SI{1}{[E]}$ and
$\sigma_\mathrm{mix} = (r_\mathrm{rasp} + \sigma_\mathrm{mono} /
2)$. The raspberry used for this publication consists of 1400 coupling
points ($\approx 5.25$ per LB unit cell), the fluid-coupling friction
parameter is $\gamma = \SI{15}{[m]/[t]}$. A visualization of the
raspberry is given in figure~\ref{fgr:raspberry_cut}, a representative
snapshot of the readily set-up system is shown in
figure~\ref{fgr:snapshot}.

\section{Obtaining AC susceptibility spectra from simulations \label{sec:susceptometry}}

The AC susceptibility is the differential response of a system's
magnetization $M(t)$ to an oscillating external magnetic field $H(t)$,
\textit{i.e.} $\chi = \mathrm{d}M(t)/\mathrm{d}H(t)$. The result is a
complex value $\chi = \chi' - \textit{i} \chi''$.

\subsection{For reference: Debye model \label{sec:debye}}

Before introducing the methods for measuring AC susceptibility spectra
in the simulation, we briefly present the result of the theoretical
Debye model.\cite{debye29a} It was first derived by Debye (1929),
originally to describe the relaxation of the rotational polarization
of molecules, but generally captures the rotational relaxation process
of any dipole.  The model makes a few assumptions. Namely, it
describes the magnetization of non-interacting ideal dipoles that
couple to a fluid with frequency-independent dynamic viscosity
$\eta = \nu \rho$ \textit{via} rotational Stokes friction
$\zeta_\mathrm{r} = 8 \pi \eta r^3$.  The magnetic susceptibility
analytically evaluates to
\begin{equation}
\chi(\omega) = \frac{\chi_0}{1 + \mathrm{i} \omega \tau},
\end{equation}
with $\chi_0$ the zero-frequency susceptibility and $\tau$ the
characteristic decay time of the magnetic moment.

For a raspberry particle in a LB fluid, as used in our model, these
assumptions are fulfilled. We can therefore use the Debye model as a
reference curve to benchmark our measurement techniques against.  In
our model we consider magnetically blocked particles only and thus do
not allow internal (N\'eel-)relaxation of the magnetic
moment. Instead, decorrelation of the orientation of the dipole moment
only happens via (rotational) Brownian diffusion. As a result we have
\begin{equation}
\tau = \tau_\mathrm{B} = \frac{3 \eta V_\mathrm{h}}{k_\mathrm{B} T}, \label{eqn:tau_brown}
\end{equation}
with Boltzmann's constant $k_\mathrm{B}$, temperature $T$, and the
hydrodynamic volume
$V_\mathrm{h} = 4 \pi r_\mathrm{h}^3 / 3$.\cite{debye29a}

\subsection{Direct measurement \label{sec:direct}}

The straightforward approach of measuring the AC susceptibility of the
simulation is by faithfully reproducing the experimental measurement
technique \textit{i.e.} measuring per frequency $f$: for any chosen
angular frequency $\omega = 2 \pi f$, a time-dependent alternating
magnetic field
\begin{equation}
	\vec{H}(t) = \vec{H}_0 \sin(\omega t),
\end{equation}
with amplitude $\vec{H}_0$ is explicitly applied to the simulation
box, causing a torque on the magnetic particle. The magnetization
response of the system is measured \textit{via} the component of the
particle's magnetic moment that points into field direction.

\begin{figure}[h]
	\centering
	\includegraphics[width=\linewidth]{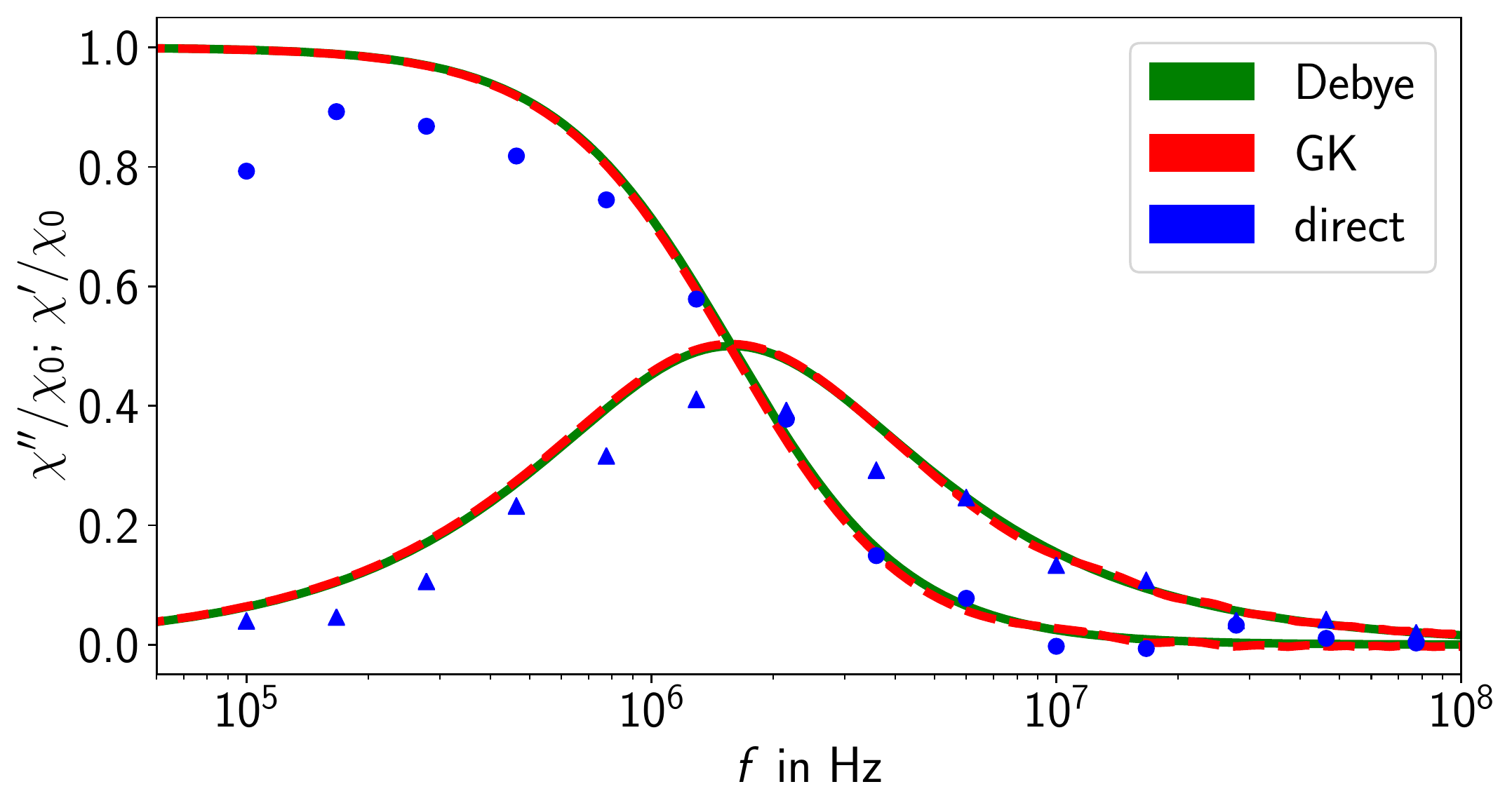}
	\caption{AC susceptibility spectra measured from the
          simulation of a MNP in a Newtonian fluid without any
          polymers. Blue data points are results of direct
          measurements (Langevin parameter $\alpha = 2$). For
          reference, the theoretical Debye model is shown (green).The
          spectrum obtained using the Green-Kubo approach (red)
          matches the theoretical model quite accurately.}
	\label{fgr:gk_vs_debye}
\end{figure}

For weak coupling to the external field, the magnetization
$\vec{M}(t)$ will also have a sinusoidal shape
\begin{equation}
\vec{M}(t) = \vec{M}_\mathrm{max} \sin(\omega t - \vartheta), \label{eqn:direct_response}
\end{equation}
but with a frequency-dependent phase shift $\vartheta$. However, this
signal will be superimposed by thermal noise. To bring out the signal,
averaging over several periods is necessary, as will be discussed
shortly.
\begin{figure}[h]
	\centering
	\includegraphics[width=\linewidth]{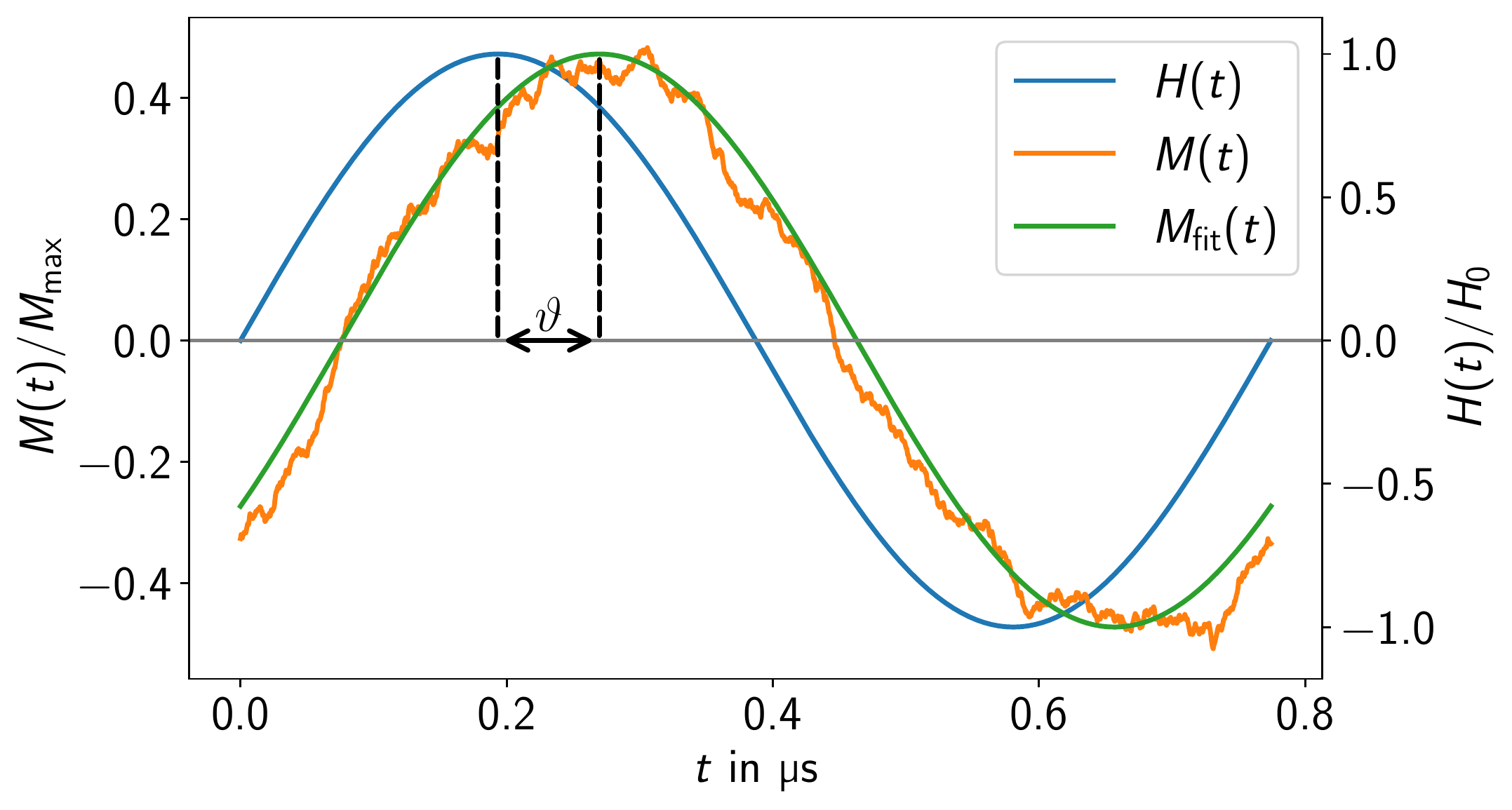}
	\caption{Direct measurement of the susceptibility: an external
          magnetic AC field $H$ (frequency $f=\SI{1291549}{\hertz}$)
          is applied to the system and its magnetization response
          $M(t)$ is measured. Due to Brownian diffusion this signal
          will be noisy and has thus to be averaged over several
          periods, 64 here. The mean magnetization (orange) is then
          fitted (green, for functional form see
          eq.~\ref{eqn:direct_response}) to obtain the phase shift
          $\theta$.}
	\label{fgr:direct_measurement}
\end{figure}
The phase shift is then used to obtain the complex magnetic
susceptibility $\chi = M(t) / H(t) = \chi' - \mathrm{i} \chi''$, where
the real part $\chi'$, that relates to reversible magnetization, is in
phase with the external field. The imaginary component $\chi''$ is
related to losses and can be computed as the out-of-phase part of the
magnetization response. We get
\begin{align}
	\chi' &= \frac{M_\mathrm{max}}{H_0} \cos\left( \vartheta \right),\\
	\chi'' &= \frac{M_\mathrm{max}}{H_0} \sin\left( \vartheta \right).
\end{align}
A visualization of the analysis is shown in
figure~\ref{fgr:direct_measurement}.  Exemplary results of such direct
measurements for a MNP in a Newtonian fluid are shown in
figure~\ref{fgr:gk_vs_debye}. For reference, the corresponding Debye
model result is also shown. The positions of imaginary susceptibility
peaks match quite well. The same is true for the real part of the
susceptibility in that region. There are, however, some deviations
from the curve predicted by Debye. This can be attributed to the
non-linear nature of the Langevin magnetization law.\cite{odenbach03a, weeber18a}  Typically, the so-called Langevin parameter is used to
characterize the magnetic interaction strength. The Langevin parameter
\begin{equation}
\alpha = \frac{\mu_0 m H}{k_\mathrm{B} T}
\end{equation}
compares the magnetic interaction energy between the particle's dipole
$m$ and the field to the thermal energy. Here, $\mu_0$ is the magnetic
constant. In order to minimize non-linear effects in the
susceptibility spectra, one would like to choose $\alpha \ll
1$. However, choosing a low $\alpha$ simultaneously increases the
relative importance of thermal fluctuations in the system -- the
signal-to-noise ratio drops significantly, which means much longer
simulation times are necessary for comparable statistics.  Minimizing
non-linear effects when using direct measurements thus comes at a high
computational cost. Another obvious shortcoming of this technique is
that each simulation yields the complex susceptibility only for the
chosen frequency $\omega$. To sample complete susceptibility spectra
with high resolution is therefore hardly feasible.  However, direct
measurements may come in handy when observing interesting behaviour at
specific frequencies.

\subsection{Green-Kubo approach \label{sec:gk}}

In this section, we discuss an approach that allows one to obtain the
full AC susceptibility spectrum from a single simulation, assuming
that the applied external magnetic field is small.  In general, the
dynamic response of a magnetic suspension to an applied magnetic field
is non-linear. However, for small fields, $\alpha \ll 1$, linear
response can be assumed. In the framework of linear response theory,
the so-called Green-Kubo relations are a general type of
fluctuation-dissipation theorems that connect physical quantities such
as polarization or complex conductivity to time correlation functions
of associated dynamic variables.\cite{kubo57a} Also for the magnetic
susceptibility one such Green-Kubo relation can be derived. It reads
\begin{equation}
\chi(\omega) = \frac{1}{V k_\mathrm{B} T} \left[ \langle \vec{M}(0) \vec{M}(0) \rangle - \mathrm{i} \omega \int_0^\infty \exp(- \mathrm{i} \omega t) \langle \vec{M}(t) \vec{M}(0) \rangle \, \mathrm{d}t \right], \label{eqn:gk_susceptibility}
\end{equation}
where $V$ is the system's volume. The integral part of this equation
is the Fourier-Laplace transform of the time. Applying this relation,
a full AC susceptibility spectrum can be directly obtained from one
single steady-state simulation (without an explicit external magnetic
field).

\begin{figure}[h]
	\centering
	\includegraphics[width=\linewidth]{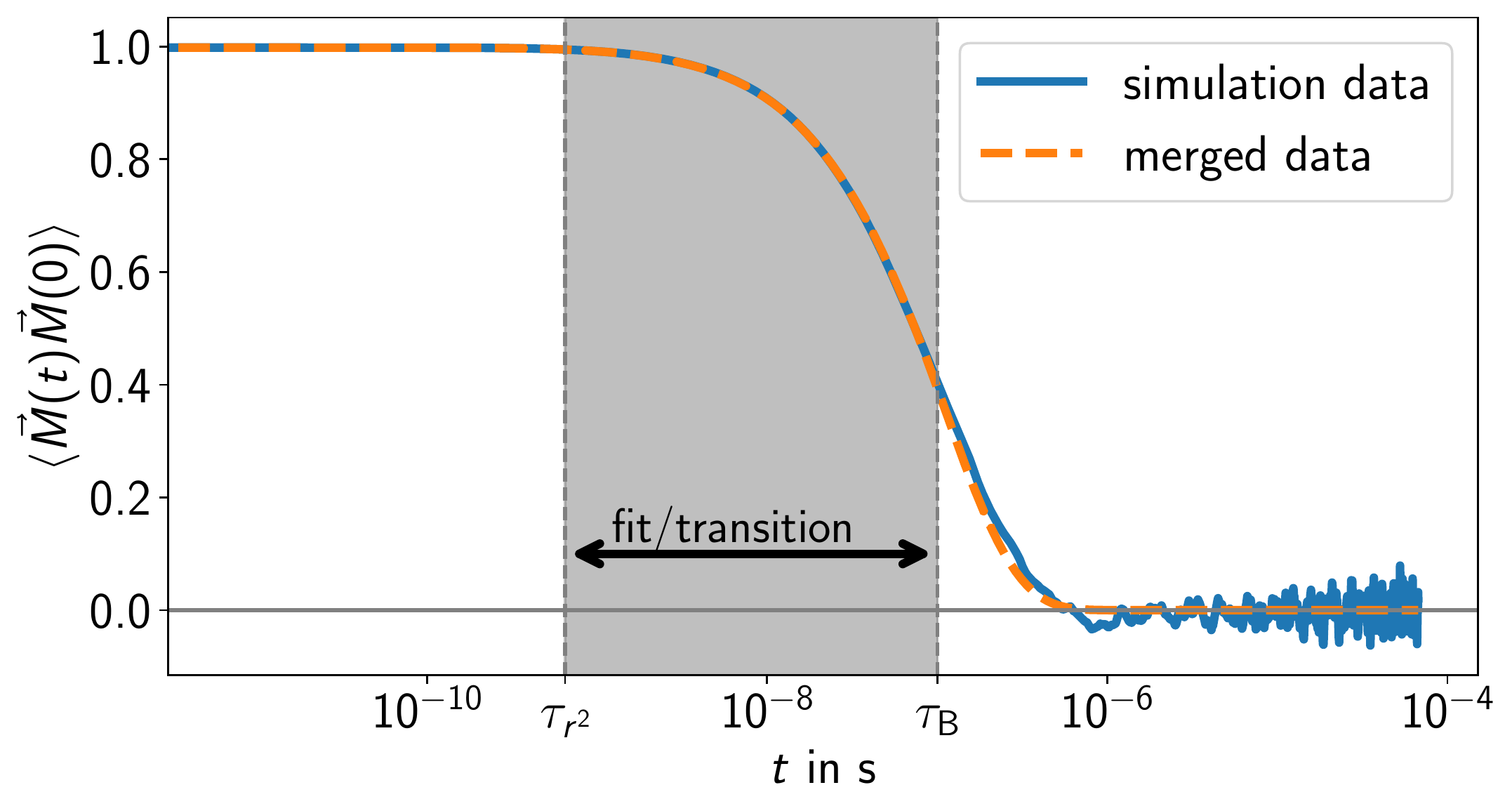}
	\caption{Representative example for the time auto-correlation
          function of the magnetization obtained from our simulation
          model (blue solid line). Here, a MNP in a Newtonian fluid
          was used. To get rid of the Brownian noise in the long-time
          tail, an analytic exponential function is fitted to the
          data, the fit region is the indicated grey area. The same
          frequency range is used to linearly transition from the numerical correlation data to the analytical tail fit (see eqs.~\ref{eqn:f_merged} and \ref{eqn:f_transition}). The resulting
          correlation function is shown by the orange dashed line.}
	\label{fgr:tapering}
\end{figure}

As this approach solely relies on the auto-correlation of the magnetic
moment as an input, one has to make sure that this quantity is well
sampled even for large lag times. As a consequence, simulation runs
have to be significantly longer than with the direct method. At this
point, the simplifications in the used parameter set (decreased
viscosity and water density for the magnetic particle) significantly
decrease computation time by speeding up the decorrelation of the
magnetic moment.  Still, in practice, dealing with the integral part
can be challenging, as thermal noise in the long-time tail of the
auto-correlation function of the magnetic moment is amplified by the
frequency.  For infinite runs noise in the long-time tail would
cancel, leaving a pure decay signal.  A common approach to mitigate
the impact of tail noise is thus to fit an analytic expression for the
tail -- a typical choice would be a simple exponential decay,
\begin{equation}
	f_\mathrm{fit}(t) = \langle \vec{M}(0) \vec{M}(0) \rangle_\mathrm{fit} \exp\left( - \frac{t}{\tau_\mathrm{fit}} \right),
\end{equation}
with $\langle \vec{M}(0) \vec{M}(0) \rangle_\mathrm{fit}$ the
zero-frequency limit of the magnetization correlation and
characteristic decay time $\tau_\mathrm{fit}$.
Note that depending on the system, more involved functional descriptions may be required to properly capture the decay behaviour, such as the expression for an isolated particle in a viscoelastic medium derived by \citet{ilg18a}. Deviations from a pure exponential decay behaviour are also expected for interacting and polydisperse ferrofluid systems.\cite{ivanov18a}
For our setup, the decay behaviour of the long-time tail is actually well captured using the simple exponential decay.
For the well-sampled low-frequency region the numerical data is kept, whereas the fitted function is used for the long-time tail.
Between these two regions a smooth transition is made over a certain frequency range.
The merged data is given by
\begin{equation}
	\langle \vec{M}(t) \vec{M}(0) \rangle = f_\mathrm{t}(t) \langle \vec{M}(t) \vec{M}(0) \rangle_\mathrm{num} + [1-f_\mathrm{t}(t)] f_\mathrm{fit}(t), \label{eqn:f_merged}
\end{equation}
where $f_\mathrm{t}(t)$ is a linear transition function over the interval $[\tau_\mathrm{l},\tau_\mathrm{u}]$,
\begin{align}
	f_\mathrm{t}(t) =
	\begin{cases}
		1 & t \leq \tau_\mathrm{l}, \\
		(\tau_\mathrm{u} - t) / (\tau_\mathrm{u} - \tau_\mathrm{l}) & \tau_\mathrm{l} < t < \tau_\mathrm{u}, \\
		0 & t \geq \tau_\mathrm{u}. \label{eqn:f_transition}
	\end{cases}
\end{align}
For a visualization of this procedure see figure~\ref{fgr:tapering}.
Here, we decided to choose the same time range for fitting and the linear transition between numerical data and analytical tail.
As lower boundary ($\tau_\mathrm{l}$), we use the viscous time of our system $\tau_{r^2} = \rho_\mathrm{w} r_\mathrm{rasp}^2 / \eta = \SI{0.58}{\nano\second}$ (the time it takes fluid momentum to diffuse by one colloidal radius).\cite{fischer15a, zwanzig75a} As upper limit ($\tau_\mathrm{u}$), we choose the Brownian relaxation time of the Newtonian system $\tau_\mathrm{B} = \SI{100.8}{\nano\second}$, see equation~\ref{eqn:tau_brown}.

The integral in eq.~\ref{eqn:gk_susceptibility} can then be
efficiently computed \emph{via} the Fourier-Laplace-transform of resulting auto-correlation function of the magnetization.  The result of
such a Green-Kubo susceptibility measurement on a magnetic particle in
a Newtonian fluid is shown in figure~\ref{fgr:gk_vs_debye}. It
perfectly matches the expected theoretical Debye relaxation reference
curve.

\section{Results and discussion \label{sec:results}}

Using our modeling approach (section~\ref{sec:model}) with the
parameters of section~\ref{sec:parameters} -- based on the
experimental system of \citet{roeben14a} -- we set up simulations to
study the signatures that changes to the polymeric surroundings
generate in the measured susceptibility spectra. Here, we will focus
on the effects of (1) varying the polymer volume fraction and (2) the
polymer chain length.

To obtain full AC susceptibility spectra, we use the Green--Kubo
approach described in section~\ref{sec:gk}, which by default operates
in the linear response limit.  The linear response assumption also
holds in the experiment as we can see from the estimated experimental
Langevin parameter.  With field amplitude
$H_{0, \mathrm{exp}} =
\SI{0.4}{\kilo\ampere\per\meter}$,\cite{roeben14a} temperature
$T_\mathrm{exp} = \SI{25}{\celsius}$, and magnetic moment of the
particle $m = \SI{3.6e-19}{\ampere\meter\squared}$,\cite{hess19a} we
obtain
\begin{equation}
	\alpha_\mathrm{exp} =  \frac{m \mu_0 H_{0, \mathrm{exp}}}{k_\mathrm{B} T_\mathrm{exp}} \approx 0.044 \ll 1,
\end{equation}
which confirms that the experiments do indeed take place in the
linear response limit. For magnetically blocked particles the magnetic
moment scales with the particles' core volume,
$m \sim r_\mathrm{rasp}^3$ (assuming constant thickness of surface
coating if present). As the used value was actually measured for a
\ce{CoFe2O4} particle with hydrodynamic radius
$r_\mathrm{h} = \SI{10.7}{\nano\meter}$ (instead of our
$r_\mathrm{h} = \SI{7.2}{\nano\meter}$),\cite{hess19a} we even
overestimate the Langevin parameter $\alpha_\mathrm{exp}$, here.

With these parameters, we also estimate the relative strength of
inter-particle dipolar interactions using the dipolar interaction
parameter
\begin{equation}
	\lambda = \frac{\mu_0 m^2}{4 \pi (2 r_\mathrm{rasp})^3 k_\mathrm{B} T},
\end{equation}
which compares the per-particle dipolar energy for two touching
particles in head-to-tail configuration to the thermal energy. We
obtain $\lambda_\mathrm{exp} \approx 0.25$, a value that indicates
rather small dipolar interactions (for particles to show notable
structure, one would expect a $\lambda$ larger than $\sim 2$).
Particle--particle interactions become important for sufficiently high
densities of MNPs and strength of dipolar interactions.\cite{ivanov18a, sindt16a,
  ivanov01a, mendelev04a} Here, however, the low dipolar interaction
parameter and volume fraction ($<\SI{1}{\percent}$) imply that
particle--particle interactions do not significantly influence the
susceptibility. This justifies our approach of studying a single
magnetic particle in its surrounding.

\begin{figure}[h]
	\centering
	\includegraphics[width=\linewidth]{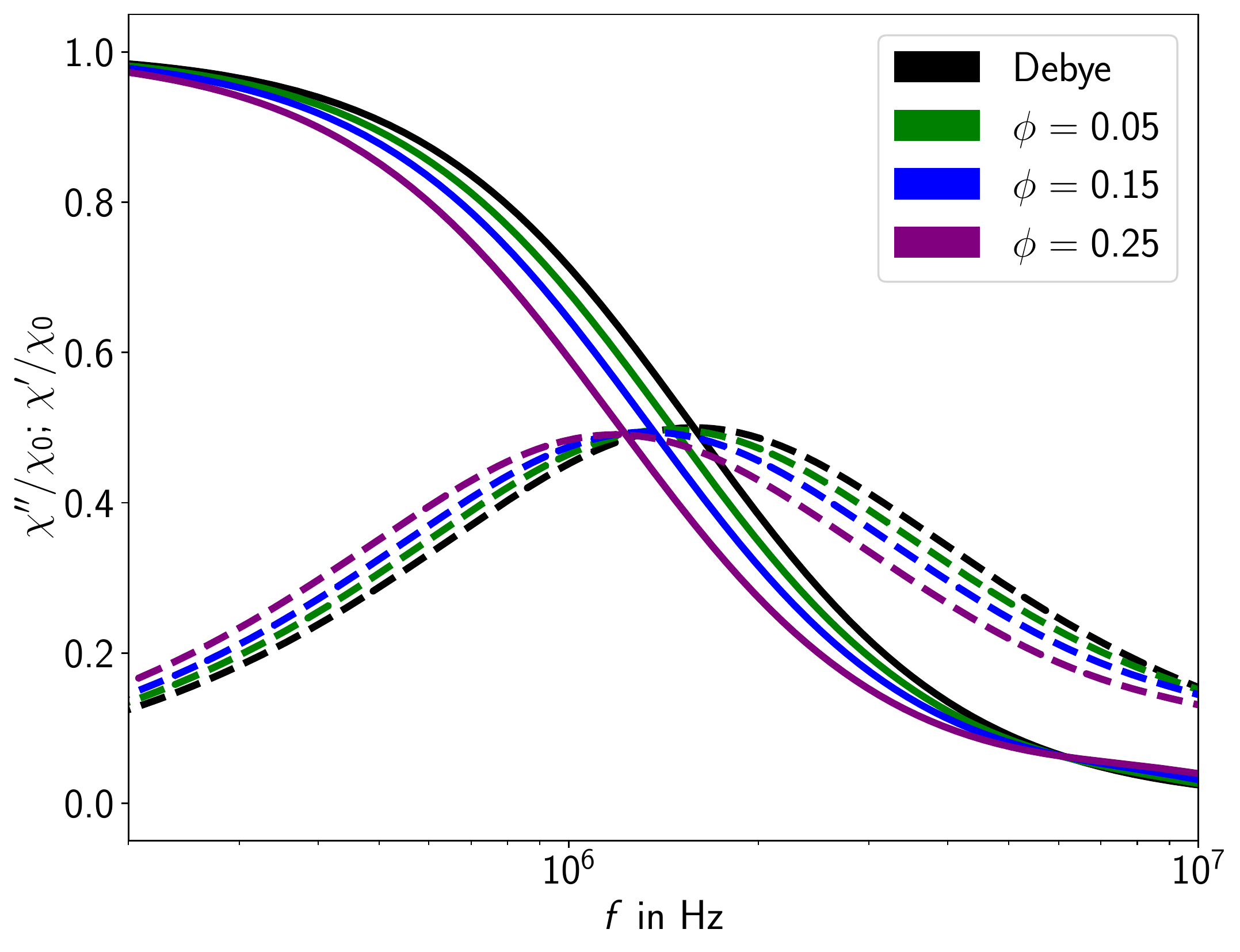}
	\caption{The AC susceptibility spectra obtained from
          simulations with polymers of chain length $N = 15$ at
          different volume fractions. Increasing the polymer volume
          fraction leads to a shift towards lower frequencies. Solid
          lines indicate the real, dashed lines the imaginary part of
          the susceptibility.}
	\label{fgr:varying_v_fraction}
\end{figure}

We start with simulations of varying polymer volume fraction.  The
polymer volume fraction refers only to the polymer solution,
\textit{i.e.}, it ignores the volume of the magnetic particles. In
general, the polymer volume fraction is thus given by
\begin{equation}
	\phi = \frac{N_\mathrm{poly} N \left(\frac{\sigma_\mathrm{mono}}{2}\right)^3}{\frac{3}{4 \pi} V_\mathrm{box} - N_\mathrm{rasp} r_\mathrm{rasp}^3},
\end{equation} 
with number of polymers $N_\mathbf{poly}$, monomer units per polymer
$N$, and the radius per monomer unit $\sigma$ (to
calculate its excluded volume). The box volume is given by
$V_\mathrm{box}$, $N_\mathrm{rasp}$ is the number of raspberries with
radius $r_\mathrm{rasp}$ inside the box. For the parameters used in
our simulations, see section~\ref{sec:parameters}.  To study the
effect of increasing polymer concentrations, we choose a constant
chain length of $N = 15$ and systematically vary the
volume fraction $\phi \in \{0.05,\,0.15,\,0.25\}$.
The magnetic moment of the
particle is recorded every $0.1$ time unit (every
\SI{6.7}{\pico\second}).  The total number of samples per simulations
run is $N_\mathrm{samples} = \SI{2e7}{}$. The resulting susceptibility
spectra are shown in figure~\ref{fgr:varying_v_fraction}.
With increasing polymer fraction, the susceptibility curves experience
a shift towards lower frequencies.  As the Debye shape of the curve is
preserved for all cases, these shifts signify an increase of the
effective viscosity around the MNP. Even at the highest
concentration, no entanglement of polymers is expected -- as will be
discussed shortly, both polymer length and volume fraction would have to be considerably
higher.

\begin{figure}[h]
	\centering
	\includegraphics[width=\linewidth]{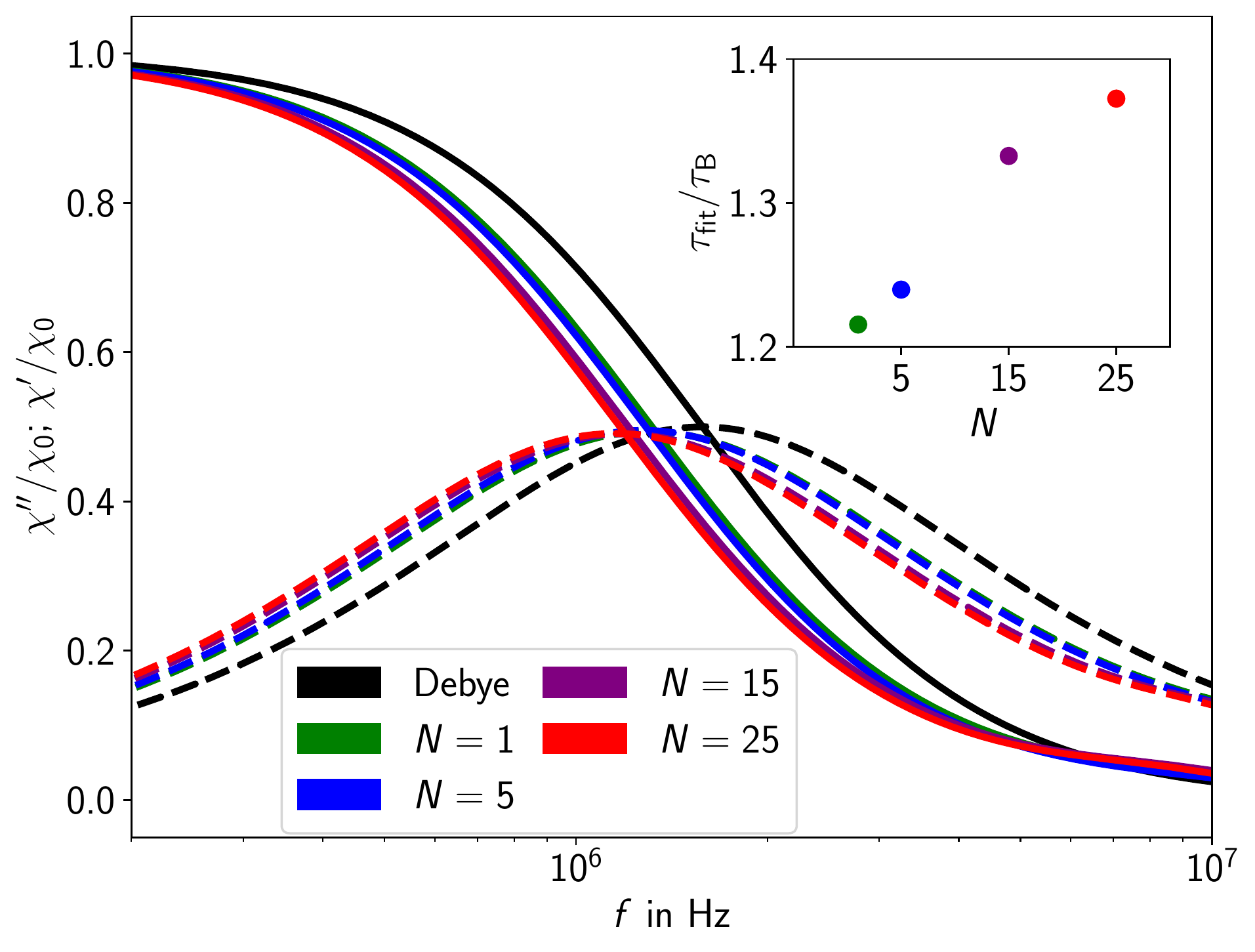}
	\caption{AC susceptibility spectra for magnetic particles in
          solution with volume fraction of polymers $\phi = 0.25$. The
          chain length of the polymers $N$ is varied. The longer the
          chains are, the more the spectra get shifted towards lower
          frequencies. Solid lines indicate the real, dashed lines the
          imaginary part of the susceptibility. The inset shows the corresponding fitted relaxation times $\tau_\mathrm{fit}$.}
	\label{fgr:vaying_chain_length}
\end{figure}

Let us now turn to the simulations at a constant volume fraction
$\phi = 0.25$ where the polymer chain length is systematically varied
$N \in \{1,\,5,\,15,\,25\}$. Results are shown in
figure~\ref{fgr:vaying_chain_length}. Again, a shift of the measured
spectra towards lower frequencies is observed while maintaining the
Debye shape, where larger shifts occur with increasing chain length.

It can be seen from the figure, that already the addition of
unconnected monomers causes a significant shift in frequency.
In our computational model, the 
monomer beads are coupled to the fluid \textit{via} the thermalized
friction force of equation~\ref{eqn:coupling_force}, thus slowing down
fluid dynamics.  This registers as an increased effective fluid
viscosity, causing a higher relaxation time for the systems'
magnetization.  An overview of the relaxation times obtained \textit{via}
fitting the long-time tail (section~\ref{sec:gk}) can be found in table \ref{tbl:decay_times}.
\begin{table}[h]
	\small
	\caption{Overview of the zero-frequency magnetization
          correlations and decay times fitted using the
          procedure described in section~\ref{sec:gk}, for simulations at
          different polymer volume fraction $\phi$ and number of
          monomers per chain $N$.}
	\label{tbl:decay_times}
	\begin{tabular*}{0.48\textwidth}{@{\extracolsep{\fill}}lrll}
          \hline
          $\phi$ & $N$ & $\langle \vec{M}(0) \vec{M}(0) \rangle_\mathrm{fit}$ & $\tau_\mathrm{fit} / \tau_\mathrm{B}$ \\
          \hline
          0.00 & 0 & 1.00 & 1.010 \\
          0.05 & 15 & 0.99 & 1.089 \\
          0.15 & 15 & 0.99 & 1.185 \\
          0.25 & 1 & 0.99 & 1.216  \\
          0.25 & 5 & 0.99 & 1.240 \\
          0.25 & 15 & 0.98 & 1.333 \\
          0.25 & 25 & 0.98 & 1.373 \\	
		\hline
	\end{tabular*}
\end{table}
By increasing both, the length of polymers and their concentration, this purely hydrodynamic effect is intensified through the increased coupling to the polymer matrix.
The observed behaviour matches the experimental observations of
\citet{roeben14a}  In the experiment, PEG solutions with different
chain lengths and polymer concentrations were studied.  As in our
simulation results, for low polymer fractions (depending on the chain
length) the experimentally measured susceptibility spectra are simple
Debye-type curves with a single relaxation time, which directly depends on the effective viscosity of the polymer suspension.
With increasing PEG
concentration and chain length, however, the experimentally measured
spectra broaden and some measurements also suggest the existence of a
second relaxation process.\cite{roeben14a} The emerging deviation from
Newtonian behaviour is most noticeable above the overlap concentration
of the respective solution. This indicates entanglement of polymer
chains as a main cause.  The broadening of Debye spectra for spherical
MNPs in increasingly elastic environments was also observed by other
experimental groups.\cite{remmer17a}

To get a feeling for the polymer lengths required to observe
entanglement in the simulation model, the work of \citet{kremer90a} on
polymer melts may serve as an indication.  Using a \emph{FENE}
potential (instead of harmonic) to connect the MD beads within
polymers and a slightly different parameter set, they found an
entanglement length of $N_\mathrm{e} = 35$ at a volume fraction of $\phi = 0.45$.\cite{kremer90a}
In our simulations we are well below these
values, with maximum $N = 25$ and $\phi = 0.25$. Despite the deviations in modelling, we thus expect that polymer chains would have to be significantly longer and used at higher volume fractions before seeing effects of entanglement.

As already described, our model makes several simplifying assumptions such as the use of coarse-grained polymers and using one tenth the viscosity of water for our fluid.
Our results are therefore not expected to quantitatively reproduce the experimental findings. To get an idea of the relative strength of the effects in the experiment and simulation, let us make a rough comparison, neglecting polymer entanglement effects.
In the simulations we found that the use of polymers of length $N = 25$ at a polymer volume fraction of $\phi=0.25$ leads to an increase of the relaxation time by $\sim \SI{37}{\percent}$.
Experimentally, the use of \SI{25}{\text{vol.-}\percent} PEG increased the dynamic viscosity and thus the relaxation time (see eq.~\ref{eqn:tau_brown}, we are in the Debye-like Newtonian limit) by roughly a factor of $2$ for triethylene glycol and about a factor of $3$ for PEG1000.\cite{roeben14a} The effect observed in the simulations is thus roughly one order of magnitude smaller than in the experiment, possibly due to the small length of our polymers (compared to PEG1000), and the neglect of any specific interactions of the polymers with the MNPs.

As a key result, our simulations show the importance of hydrodynamics
for magnetic composite systems.  Since all pair interactions between
the MNP and surrounding monomer beads just affect the particle's
translational behaviour, hydrodynamic interactions are the only
possible source of rotational coupling.  Our simulations demonstrate
that this is already sufficient to qualitatively reproduce the
experimentally observed shifts of the AC susceptibility spectra
towards lower frequencies for both, higher polymer volume fractions
and longer chains, respectively.

\section{Conclusions}
In this work, we presented a simulation model for magnetic
nanoparticles in polymer suspensions and showed how
frequency-dependent AC susceptibility spectra can be obtained from the
simulations.  Our modelling approach is based on molecular dynamics,
coupled to an efficient lattice-Boltzmann hydrodynamics solver.  While
still costly, we managed to bring the computational complexity down to
a feasible level. To do so, we made some simplifying assumptions, such
as a certain level of coarse-graining regarding the polymers.  Using
our model, we ran simulations with parameters based on the
experimental system of \citet{roeben14a}, to study the signatures that
changes to the polymeric surroundings produce in the resulting
susceptibility spectra.  For this, we simulated a single particle in
its polymeric environment (using periodic boundary conditions).  This
is justified because of the low concentration of particles as well as
small dipolar interaction strength in the experimental system.  We
presented two approaches to obtain susceptibility spectra from our
simulations. Using direct measurement one may obtain the
susceptibility for a specific frequency, which, however, is quite
costly when it comes to sampling entire spectra over a large range of
frequencies.  We therefore introduced a second approach based on
Green-Kubo linear response theory.  To benchmark this method with our
model, we showed that the simulation of a particle in a Newtonian
fluid reproduces the curve expected from Debye relaxation theory.  We
used the Green-Kubo approach throughout our simulation study of a
spherical nanoparticle in a polymer suspension.  Matching the
experimental observations,\cite{roeben14a} we found that the AC
susceptibility spectra are shifted towards lower
frequencies,\cite{roeben14a} when either increasing the polymer volume
fraction or the polymer chain length for a fixed polymer
concentration.  In our model, hydrodynamic interactions provide the
sole source of direct coupling between the polymeric environment and
the rotational behaviour of the particle, \textit{i.e.} the Brownian
relaxation of the systems' magnetic moment.  Finding that this is
already sufficient to qualitatively reproduce the experimental trends
highlights the essential role of hydrodynamics for the behaviour of
spherical magnetic nanoparticles in complex environments.

\section*{Conflicts of interest}
There are no conflicts to declare.

\section*{Acknowledgements}
The authors acknowledge financial support by the Deutsche
Forschungsgemeinschaft (DFG) through the priority program SPP 1681
(Grant HO\,1108/23-3).



\balance


\bibliography{bibtex/icp} 
\bibliographystyle{rsc} 

\end{document}